\documentstyle{l-aa}
\input epsf
\begin{document}
\include{astro}
\renewcommand{\footnoterule}{}
\thesaurus{03(03.20.7,08.06.3)}
\title {{\em Research Note :} A critique of disentangling as a method
        of deriving spectroscopic orbits} \author{R.I. Hynes \and
        P.F.L. Maxted \thanks{ \protect\raggedright
	{\it Present address:} Astronomy Group, University 
	of Southampton, Highfield, Southampton, SO17 1BJ,
        England}} 
	\institute{Astronomy Centre, University of Sussex,
        Falmer, Brighton, BN1 9QH, England}
\date{Received ; accepted}
\offprints{\protect\raggedright R.I. Hynes [Email: rih@star.cpes.susx.ac.uk]}
\maketitle 
\markboth{R.I. Hynes \& P.F.L. Maxted: Disentangling as a
method of deriving spectroscopic orbits}{R.I. Hynes \& P.F.L. Maxted:
Disentangling as a method of deriving spectroscopic orbits}
%
%%%%%%%%%%%%%%%%%%%%%%%%%%%%%%%%%%%%%%%%%%%%%%%%%%%%%%%%%%%%%%%%%%%%%%%%%%%%%%%
%
% Abstract
%
\begin{abstract}
Multiple sets of synthetic spectra of OB-binary stars are used to test
the suitability of disentangling for deriving accurate spectroscopic
orbits. Given a set of spectra with broad phase coverage and
sufficient total integration time (almost independent of the number of
spectra), it appears that disentangling yields accurate and reliable
semi-amplitudes of the spectroscopic orbits (other parameters being
fixed).  Methods for estimating the uncertainties on the derived
semi-amplitudes are investigated.

\end{abstract}
%
%%%%%%%%%%%%%%%%%%%%%%%%%%%%%%%%%%%%%%%%%%%%%%%%%%%%%%%%%%%%%%%%%%%%%%%%%%%%%%%
%
% Introduction
%
\section{Introduction}
The disentangling technique was first presented by Simon \& Sturm
(1994, hereafter SS94) and applied to the early B-type binary, V453
Cyg.  It has subsequently been used on the O-type systems DH~Cep
(Sturm \& Simon 1994) and Y~Cyg (Simon et al.\ 1994). The original
motivation for the method was that it enabled the separation of
closely blended spectra, which were then suitable for quantitative
analysis.  Accurate separation of the spectra requires accurate
parameters for the spectroscopic orbit. Those parameters not fixed by
other information can, in principal, be determined by finding the
values of these free parameters that minimize the residual of the fit
to the observed spectra when the separated spectra are recombined.

In this research note we summarize the results of Hynes (1996), in
which multiple sets of synthetic spectra were used to investigate
whether disentangling is a viable method for determining accurate
spectroscopic orbits for a selection of OB eclipsing binary stars.
%
%%%%%%%%%%%%%%%%%%%%%%%%%%%%%%%%%%%%%%%%%%%%%%%%%%%%%%%%%%%%%%%%%%%%%%%%%%%%%%%
%
% The synthetic data
%
\section{The synthetic data}
The synthetic binary star spectra are based on a grid of single star
spectra kindly provided by Dr C.S. Jeffrey (1996). These cover the
spectral range 3600-5000\,\AA.  Each component spectrum was obtained
by linear interpolation on the single star grid using the best
estimates of $T_{\rm eff}$ and $\log g$, followed by convolution with
the appropriate rotational broadening profile. A Doppler shift was
then applied to each component based on the adopted spectroscopic
orbit and the phase desired and the two shifted spectra were added
according to luminosity ratio.  Poisson noise was applied to the
spectra to produce a realistic signal-to-noise ratio (S/N) (typically
S/N=100--200). Ten data sets (each of 15--20 spectra) were generated
for each case studied, with phases randomly chosen from a uniform
distribution excluding eclipses.  Parameters of the systems discussed
here are given in Table~\ref{ParamTable}.

\begin{table}
\caption[]{Adopted parameters of the systems studied together with
           mean velocities (and errors on this mean) derived by
           disentangling 10 sets each of 20 synthetic spectra.  The
           final column shows the population standard deviation of the
           10 sets and is a measure of the intrinsic scatter of the
           method.}
\label{ParamTable}
\begin{flushleft}
\begin{tabular}{lccccc}
\hline
\noalign{\smallskip}
&$L_2/L_1$ &$v\sin i$&\multicolumn{2}{c}{$K ({\rm km\,s}^{-1})$}& Std.
dev.\\
\cline{4-5}
     &          &(km/s)&Adopted & Derived & (km/s)\\
\noalign{\smallskip}
\hline
\noalign{\smallskip} 
A & 0.885 & 165 & 262.3 & 262.02$\pm$0.17 & 1.0 \\ 
       &       & 165 & 278.0 & 278.12$\pm$0.37 & 1.3 \\ 
\noalign{\smallskip}
B & 0.875 & 100 & 235.0 & 235.08$\pm$0.47 & 1.4 \\ 
  &       & 100 & 245.0 & 245.04$\pm$0.29 & 0.9 \\ 
\noalign{\smallskip}
C & 1.156 & 165 & 223.0 & 223.24$\pm$0.27 & 0.8 \\ 
  &       & 200 & 205.0 & 204.99$\pm$0.26 & 0.8 \\ 
\noalign{\smallskip}
D & 0.195 & 165 & 135.0 & 135.09$\pm$0.17 & 0.5 \\ 
  &       & 140 & 190.0 & 189.83$\pm$0.95 & 2.8 \\ 
\noalign{\smallskip}
E & 0.753 &  \hspace*{0.55em}85 & 145.1 & 145.08$\pm$0.08 & 0.2 \\ 
  &       &  \hspace*{0.55em}85 & 145.8 & 145.78$\pm$0.16 & 0.5 \\ 
\noalign{\smallskip}
F & 0.868 & 200 & 100.0 & 102.45$\pm$1.33 & 4.8 \\ 
  &       & 200 & 110.0 & 108.02$\pm$1.34 & 4.6 \\ 
\noalign{\smallskip}
\hline
\end{tabular}
\end{flushleft}
\end{table}

For eclipsing systems parameters other than the semi-amplitudes of the
spectroscopic orbit ($K_1$ and $K_2$) can usually be determined from
photometric data.  We can therefore estimate optimum values of $K_1$
and $K_2$ for each of the ten synthetic data sets using a robust grid
searching method. Briefly, the residual, $r$ is calculated for a range
of $K_1$ and $K_2$ (using our own implementation of SS94's algorithm),
using successively finer grids centered on the maximum of the previous
grid. The mean difference between the ten resulting $K_1$, $K_2$
values and the adopted values indicates the reliability of the method
(i.e.\ this is a test for systematic errors) and the population
standard deviation of the ten sets measures its accuracy.
%
%%%%%%%%%%%%%%%%%%%%%%%%%%%%%%%%%%%%%%%%%%%%%%%%%%%%%%%%%%%%%%%%%%%%%%%%%%%%%%%
%
% Phase distribution
%
\section{Phase distribution}
Existing data sets for binary stars are often concentrated around
phases near quadrature to minimize problems with blending inherent
with established techniques for measuring radial velocities.  This is
not the best approach to take with the disentangling procedure.  It is
clear from the way the equations are set up that the system is best
conditioned if the phases cover the full velocity range.  This is
confirmed by numerical tests (Hynes 1996).  Thus it appears that, in
contrast to other techniques, the disentangling process will work best
when given as broad a phase coverage as possible.  This will often be
to the observer's advantage. 
%
%%%%%%%%%%%%%%%%%%%%%%%%%%%%%%%%%%%%%%%%%%%%%%%%%%%%%%%%%%%%%%%%%%%%%%%%%%%%%%%
%
% Integration time
%
\section{Integration time}
\begin{figure}
%\picplace{6cm}
\epsfbox{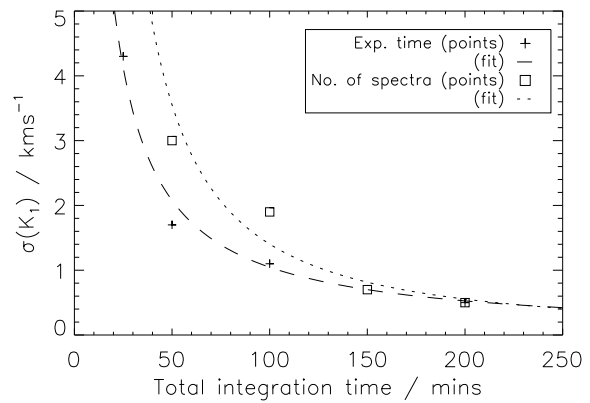}
\caption[]{The effects of varying the exposure time and number of
           spectra on the uncertainty in the deduced primary
           semi-amplitude, $K_1$, Star~C.  Similar results are
           obtained for $K_2$.}
\label{KFigure}
\end{figure}

We next consider the effects of varying the number of spectra obtained
and the duration of each exposure.  Both of these can be thought of as
varying the total integration time on the target.  The combinations
chosen are listed in Table~\ref{IntTimeTable}, together with the
resulting uncertainties in deduced parameters, again, for star C.
They are plotted in Fig.~\ref{KFigure} as a function of total
integration time for the primary velocity; similar curves are obtained
for $K_2$.  Two power law best fits are plotted; one for variations in
number of spectra and the other for variations in exposure time.

\begin{table}
\caption[]{Effect of varying the number of spectra and the integration time
           for Star C.}
\label{IntTimeTable}
\begin{flushleft}
\begin{tabular}{cccc}
\hline
\noalign{\smallskip}
Number     & Exposure & $\sigma (K_{1})$ & $\sigma (K_{2})$ \\
of spectra & time (s) & (km\,s$^{-1}$)     & (km\,s$^{-1}$)     \\ 
\noalign{\smallskip}
\hline
\noalign{\smallskip}
20                & 600      & 0.5              & 0.8              \\
15                & 600      & 0.7              & 0.7              \\
10                & 600      & 1.9              & 1.0              \\
\hspace*{0.55em}5 & 600      & 3.0              & 2.0              \\
\noalign{\smallskip}
\hline
\noalign{\smallskip}
20                & 600      & 0.5              & 0.8              \\
20                & 300      & 1.1              & 1.2              \\
20                & 150      & 1.7              & 1.6              \\
20 & \hspace*{0.55em}75      & 4.3              & 4.6              \\
\noalign{\smallskip}
\hline
\end{tabular}
\end{flushleft}
\end{table}

It is clear from this figure that there is no significant difference
between the two curves i.e.\ it matters little whether we change the
number of spectra obtained or the exposure time of each; it is the
{\em total} exposure time that matters.
%
%%%%%%%%%%%%%%%%%%%%%%%%%%%%%%%%%%%%%%%%%%%%%%%%%%%%%%%%%%%%%%%%%%%%%%%%%%%%%%%
%
% Accuracy and reliability of the derived orbits
%
\section{Accuracy and reliability of the derived orbits}
In Table~\ref{ParamTable} we show the adopted $K_1$, $K_2$ values and
the mean deduced values derived from 10 sets of synthetic spectra. All
the systems studied, with the exception of star F, show no systematic
difference between the adopted and deduced $K_1$, $K_2$ values i.e.\
disentangling yields reliable orbital parameters for systems with
$v\sin i \la K$. The accuracy of disentangling is estimated from the
standard deviation of $K_1$, $K_2$ between the 10 sets
(Table~\ref{ParamTable}, final column).  Accuracies $\sim 1\%$ are
found with the exception of star F and star D, presumably due to the
high rotational velocity and extreme luminosity ratio, respectively.
We also looked at the effect of rectification errors $\sim 5\%$ on the
accuracy and reliability of deduced $K_1$, $K_2$ values but found no
significant effect.
%
%%%%%%%%%%%%%%%%%%%%%%%%%%%%%%%%%%%%%%%%%%%%%%%%%%%%%%%%%%%%%%%%%%%%%%%%%%%%%%%
%
% Estimation of errors
% 
\section{Estimation of errors} 
In practice, given a single real data set, we require some method to
estimate the accuracy of the $K_1,K_2$ values derived from
disentangling. In this section we test two proposed methods.
%
% Curvature analysis
%
\subsection{Curvature analysis}
\label{CurvatureSection}
As outlined by SS94 and elaborated by Sturm (1994), we can estimate
the uncertainties in the derived parameters by converting $r$ to
$\chi^2$ using the mean S/N of the observed spectra.  The curvature of
the $\chi^2$ surface (as a function of the unknown parameters) then
yields the uncertainties in the free parameters.  The most general
method of obtaining confidence intervals from this surface is to
construct a contour of constant $\chi^2$, defined by
$\chi^2=\chi^2_{\rm min}+\Delta\chi^2$ and {\em project} this onto the
$K_1$ and $K_2$ axes.  For the $1\sigma$, two parameter case
$\Delta\chi^2=2.3$.  This assumes a normal distribution of errors
which is a good approximation for Poisson noise in the high
signal-to-noise limit.

We determined this contour for star E using 10 sets of 20 spectra. The
confidence limits on $K_1$ and $K_2$ are shown in
Table~\ref{ChiTable}. While the deduced uncertainties for $K_1$ are
satisfactory, those for $K_2$ are somewhat low.  This is not unique to
this data set -- it is a common problem.

\begin{table}
\caption[]{$1 \sigma $ confidence limits from the projection of the
           $\chi^2 = \chi_{\rm min}^2 + 2.3$ contour for star E. The
           population standard deviations between the data sets in
           this table are 0.25 for $K_{1}$ and 0.51 for $K_{2}$.}
\label{ChiTable}
\begin{flushleft}
\begin{tabular}{lcc}
\hline
\noalign{\smallskip}
Data & $K_{1}$        & $K_{2}$        \\
set  & (km\,s$^{-1}$) & (km\,s$^{-1}$) \\ 
\noalign{\smallskip}
\hline
\noalign{\smallskip} 
1    & $144.8\pm0.2$ & $145.3\pm0.4$ \\
2    & $145.0\pm0.3$ & $145.6\pm0.4$ \\
3    & $145.1\pm0.2$ & $145.8\pm0.4$ \\
4    & $144.9\pm0.3$ & $145.2\pm0.4$ \\
5    & $145.1\pm0.3$ & $146.9\pm0.3$ \\
6    & $145.5\pm0.3$ & $145.6\pm0.3$ \\
7    & $145.2\pm0.3$ & $145.7\pm0.3$ \\
8    & $144.7\pm0.3$ & $145.7\pm0.6$ \\
9    & $145.4\pm0.3$ & $145.6\pm0.4$ \\
10   & $145.1\pm0.2$ & $146.4\pm0.4$ \\
\noalign{\smallskip} 
\hline
\end{tabular}
\end{flushleft}
\end{table}

This is not quite the approach taken by SS94 who instead measure the
curvature of the surface at the minimum, represented by the curvature
matrix, $\Gamma$.  The confidence intervals, $\delta K_i$, are then
given by $\delta K_i=\pm\sqrt{\Delta\chi^2}\sqrt{(\Gamma^{-1})_{ii}}$
(Sturm 1994, Press et al.\ 1992). We determine the curvature
numerically using the data from the final grid for which points are
separated by 0.1\,km\,s$^{-1}$, which is similar to the expected
uncertainty i.e.\ a suitable value for determination of the curvature.
Provided that the surface is smooth, the resulting uncertainties are
comparable to those obtained by projection of the confidence region.
%
% Multiple spectra method
%
\subsection{Multiple spectra method}
This study was undertaken in preparation for observations using an
echelle spectrograph. We therefore decided to investigate whether the
standard deviation of the results obtain with different orders of the
echellogram would produce reliable uncertainties. We decided upon a
simplistic approach, with each order weighted equally.  This will
degrade the results somewhat, but has the advantage that it does not
involve estimating the error from a single order and is therefore
independent of methods using the curvature matrix.  The spectral range
of our synthetic spectra was equivalent to only 4 echelle orders. For
the specific case of star E, Table~\ref{EchelleTable} shows that the
results obtained by dividing spectra into four sections, each
analyzed independently (note that this used a different data run to
that in Table~\ref{ChiTable}, so the choice of random phases and the
noise will be different). On average, this is an unbiased estimate of
uncertainties but the low number of ``independent trials'' results in
poor estimates of the uncertainty in an individual case.

\begin{table}
\caption[]{Results of disentangling ten synthetic data sets in four
           separate segments.  The population standard deviations
           between the data sets in this table are 0.4 for $K_{1}$ and
           0.7 for $K_{2}$. Those obtained by disentangling the whole
           spectra were 0.25 for $K_{1}$ and 0.48 for $K_{2}$. }
\label{EchelleTable}
\begin{flushleft}
\begin{tabular}{lcc}
\hline
\noalign{\smallskip}
Data & $K_{1}$        & $K_{2}$        \\
set  & (km\,s$^{-1}$) & (km\,s$^{-1}$) \\
\noalign{\smallskip}
\hline
\noalign{\smallskip}
1    & $145.0\pm0.3$  & $145.6\pm0.6$  \\    
2    & $145.0\pm0.3$  & $145.3\pm1.3$  \\    
3    & $145.6\pm0.4$  & $147.0\pm1.3$  \\    
4    & $144.5\pm0.2$  & $146.3\pm0.7$  \\    
5    & $145.4\pm0.3$  & $145.9\pm0.9$  \\    
6    & $144.7\pm0.2$  & $146.1\pm0.8$  \\    
7    & $145.4\pm0.3$  & $147.4\pm1.4$  \\    
8    & $145.4\pm0.2$  & $145.6\pm0.4$  \\    
9    & $145.4\pm0.5$  & $146.5\pm0.6$  \\    
10   & $145.1\pm0.3$  & $145.8\pm0.6$  \\    
\noalign{\smallskip}
\hline
\end{tabular}
\end{flushleft}
\end{table}

\begin{figure*}
%\picplace{6cm}
\epsfbox{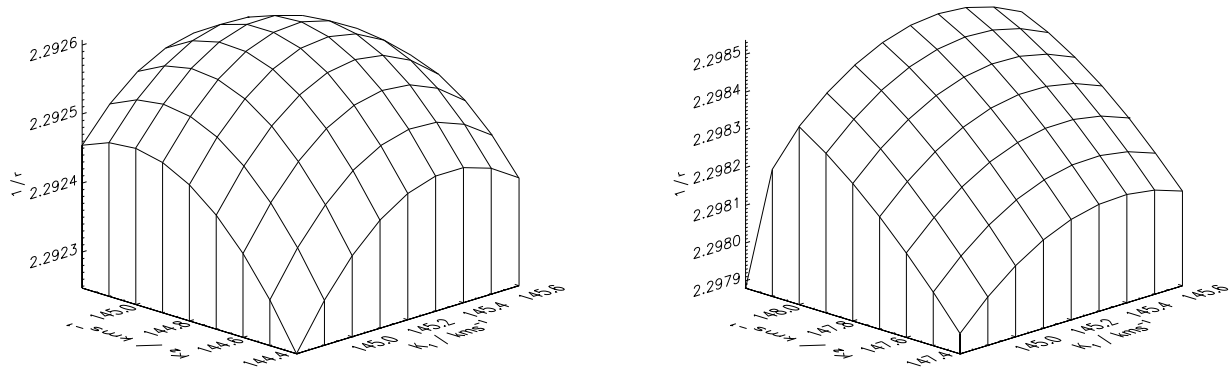}
\caption[]{Two grids resulting from disentangling synthetic spectra of
LZ~Cen differing only in their random elements.}
\label{GridFig}
\end{figure*}
%
%%%%%%%%%%%%%%%%%%%%%%%%%%%%%%%%%%%%%%%%%%%%%%%%%%%%%%%%%%%%%%%%%%%%%%%%%%%%%%%
%
% Discussion
%
\section{Discussion}
\label{Discussion}
Our experience with disentangling has led us to be cautious about
uncertainties determined from the curvature of $\chi^2$ space near its
minimum.  These rely on several assumptions which may not be valid in
the case of disentangling.  For example, it is not clear that the
disentangling process is strictly equivalent to model-fitting by least
squares minimization and that we can treat it as a two-parameter
(i.e.\ $K_1$ and $K_2$) problem. The disentangling procedure
simultaneously fits both separated component spectra and the
radial-velocity amplitudes; in effect each pixel of each separated
spectrum is a parameter to be fitted.  In applying curvature analysis
as above we are assuming that the separated spectra are simply
``nuisance parameters'' and can be ignored in the error analysis.
This may seem a reasonable assumption but the fact that curvature
analysis consistently gets the relative errors on $K_1$ and $K_2$
wrong (Section~\ref{CurvatureSection}) indicates that the r\^{o}le of
the separated spectra may be more subtle.

Statistical subtleties aside, a more pragmatic reason for caution is
shown in Fig.~\ref{GridFig}. Two grids are shown for two synthetic
data sets of the same star that differ only in their noise and phase
distribution (which are similar). One grid shows a ``ridge'' near the
minimum, a feature that was commonly seen in these grids. The
uncertainties derived from the curvature of such a surface are clearly
unreliable. In this case, constructing a surface of constant $\chi^2$
would be more appropriate.  In surfaces derived from real spectra,
even more complex structure can sometimes be seen, for example
multiple maxima or a series of parallel ridges.  Such effects will
further complicate the analysis.

The multiple spectra method appears to give unbiased estimates of the
uncertainties but requires more independent spectral sections showing
useful spectral lines than are seen in early-type stars. It might well
be applicable to late-type stars where even small spectral regions
contain a large number of features.

In attempting to analyze the accuracy of the procedure with synthetic
data sets we must consider how well we reproduce the characteristics
of real data.  One issue is that the phase sampling achieved is likely
to be less than ideal.  We have attempted to mimic this by choosing
random phases with no extra constraints (for example we do not impose
any restriction on how close in phase two observations can be -- real
observations spread over several nights could give duplicate phasing).
The sharp ridges commented on above may be one manifestation of poor
sampling.

Another way in which real data will differ is that the noise
distribution is unlikely to be the idealized uncorrelated Poisson
noise assumed in this work, e.g.\ interpolating spectra onto a
logarithmic grid is known to introduce a short scale autocorrelation
into the noise, which can lead to a ``rippling'' on the residual
surfaces.  Other effects, such as cosmic ray events and wavelength
calibration errors may also lead to noticable distortions of the
surfaces which must be identified and accounted for.  Numerical
simulations of the type discussed in this paper will be a valuable
tool in investigating these issues.  For real data however, we will
not have the luxury of obtaining multiple duplicate data sets and comparing
them.  The r\^{o}le of simulated multiple data sets is then to give us
insight into the reliability of methods such as curvature analysis.
%
%%%%%%%%%%%%%%%%%%%%%%%%%%%%%%%%%%%%%%%%%%%%%%%%%%%%%%%%%%%%%%%%%%%%%%%%%%%%%%%
%
% Acknowledgements
%
\begin{acknowledgements}
R. Hynes was supported by a PPARC Advanced Course Studentship. The
authors would like to thank Simon Jeffery for his synthetic single
star spectra and Martin Hendry for helpful discussion on matters
statistical.  P.~Maxted would like to thank Klaus Simon for his help
and advice.
\end{acknowledgements}
%
%%%%%%%%%%%%%%%%%%%%%%%%%%%%%%%%%%%%%%%%%%%%%%%%%%%%%%%%%%%%%%%%%%%%%%%%%%%%%%%
%
% References
%

%
\end{document}